\documentstyle[prd,aps]{revtex}
\begin{document}
\draft

%
%
\twocolumn[\hsize\textwidth\columnwidth\hsize\csname
@twocolumnfalse\endcsname

\preprint{SUSSEX-AST 95/8-1, astro-ph/9508003}
\title{The open universe Grishchuk--Zel'dovich effect}
\author{Juan Garc\'{\i}a-Bellido}
\address{Astronomy Centre, University of Sussex, Falmer, Brighton BN1
9QH,~~~U.~K.}
\author{Andrew R.~Liddle}
\address{Astronomy Centre, University of Sussex, Falmer, Brighton BN1
9QH,~~~U.~K.}
\author{David H.~Lyth}
\address{School of Physics and Chemistry, University of Lancaster,
Lancaster LA1 4YB,~~~U.~K.}
\author{David Wands}
\address{Astronomy Centre, University of Sussex, Falmer, Brighton BN1
9QH,~~~U.~K.}
\date{\today}
\maketitle
\begin{abstract}
The Grishchuk--Zel'dovich effect is the contribution to the microwave
background anisotropy from an extremely large scale adiabatic density
perturbation, on the standard hypothesis that this perturbation is a typical
realization of a homogeneous Gaussian random field. We analyze this effect in
open universes, corresponding to density parameter $\Omega_0<1$ with no
cosmological constant, and concentrate on the recently discussed
super-curvature modes. The effect is present in all of the low multipoles of
the anisotropy, in contrast with the $\Omega_0=1$ case where only the
quadrupole receives a contribution. However, for no value of $\Omega_0$ can a
very large scale perturbation generate a spectrum capable of matching
observations across a wide range of multipoles.  We evaluate the magnitude of
the effect coming from a given wavenumber as a function of the magnitude of
the density perturbation, conveniently specified by the mean-square curvature
perturbation. From the absence of the effect at the observed level, we find
that for $0.25\leq\Omega_0\leq 0.8$, a curvature perturbation of order unity
is permitted only for inverse wavenumbers more than one thousand times the
size of the observable universe. As $\Omega_0$ tends to one, the constraint
weakens to the flat space result that the inverse wavenumber be more than a
hundred times the size of the observable universe, whereas for $\Omega_0 <
0.25$ it becomes stronger. We explain the physical meaning of these
results, by relating them to the correlation length of the perturbation.
Finally, in an Appendix we consider the dipole anisotropy and show that it
always leads to weaker constraints.
\end{abstract}
\pacs{98.80.Cq \hspace*{4cm} Sussex preprint SUSSEX-AST 95/8-1,
astro-ph/9508003}

\vskip2pc]

\section{Introduction}

The Grishchuk--Zel'dovich effect gives the contribution to the
microwave background anisotropy from a very large scale adiabatic
density perturbation, under the standard hypothesis that this
perturbation is a typical realization of a homogeneous Gaussian random
field. In this paper we place upper bounds on the spectrum of density
perturbations on very large scales in an open universe, corresponding
to a density parameter $\Omega_0<1$ with no cosmological constant.

In order to facilitate the discussion, let us make the usual assumption
that the microwave background anisotropy, as well as the large scale
structure, arises entirely from the spectrum of density
perturbations. It is convenient to work not with the density
perturbation itself, but with the primordial curvature perturbation
that corresponds to it. On scales much smaller than the observable
universe the spectrum of the primordial curvature perturbation can be
probed directly, because there is a wealth of data relating to such
scales. At the most direct level (though of course oversimplified) one
can Fourier analyze the observed galaxy number density, and the
primordial curvature perturbation on each scale is then related to the
Fourier coefficient, up to uncertainties regarding the correct
transfer function and the possibility of biased galaxy formation. In
this way one finds that the primordial spectrum is approximately scale
independent.

On larger scales the only relevant data consist of the low multipoles
of the cosmic microwave background anisotropy, and here the situation
is less clear-cut. Roughly speaking these low multipoles measure the
low-order spatial derivatives of the curvature perturbation, averaged
over the observable universe. For this reason, the standard hypothesis
is that they probe the spectrum on scales of order the size of the
observable universe. If the spectrum is taken to be the
scale-independent extrapolation of the one measured on smaller scales,
this will indeed give a good account of the data, for either flat or
open universes.

The question is whether the spectrum on scales much larger than the
observable universe could mimic the same effect. Could the low
derivatives of the curvature perturbation, measured by the cosmic
microwave background anisotropy, come from very long wavelength and
large amplitude contributions to the spectrum? More generally, what
{\em upper limit} can one place on the long wavelength spectrum of the
curvature perturbation, by requiring that its contribution to the low
multipoles be no bigger than the observed total?

For a spatially flat universe this question was asked, and essentially
answered, by Grishchuk and Zel'dovich \cite{GZ}.  They found that a
very long wavelength contribution would be present only in the
quadrupole, and such a contribution has come to be known as the
Grishchuk--Zel'dovich effect. On the basis of upper bounds on the
quadrupole existing at that time, they showed that if the geometry
distortion is of order unity on some very large scale, then that scale
must be at least a factor of a hundred bigger than the observable
universe.

In the case of a flat universe, little has changed since their
pioneering paper. Despite the fact that the large angle anisotropies
have now been measured by the COBE satellite \cite{COBE}, there is no
indication of a contribution from very large scales affecting the
quadrupole, making it stand out relative to the higher
multipoles. Consequently, one still has only a limit on how quickly
the perturbation spectrum can rise on large scales and the numerical
value of the scale at which the curvature perturbation can reach unity
is not much changed.

Now let us turn to the Grishchuk--Zel'dovich effect in the open
universe, which necessarily explores scales on which spatial curvature
is significant barring the exceptional case that $\Omega_0$ is very
close to one. Until recently, considerable confusion existed about the
correct treatment of such scales. In order to describe a cosmological
perturbation one performs a mode expansion, using eigenfunctions of
the comoving Laplacian so that, to first-order, each mode
decouples. It has long been known to cosmologists that in the open
universe any square integrable {\em function} can be generated using
only {\em sub-curvature} modes, defined as those whose eigenvalue is
in the range $-\infty$ to $-1$ in units of the curvature
scale. Presumably with this in mind, cosmologists have retained only
these modes in the expansion. But in the cosmological application we
want to generate not a single function but a Gaussian {\em random
field}, consisting of a set of functions together with a probability
distribution, and this is done by assigning an independent Gaussian
probability distribution to each coefficient in the mode
expansion. (The spectrum is essentially the variance of the Gaussian
distribution, and it defines the random field completely.)
Cosmologists have only recently \cite{LW} discovered the fact, known
to mathematicians for half a century~\cite{krein}, that the most
general homogeneous Gaussian random field contains modes with
eigenvalue in the entire range from $-\infty$ to $0$, including not
only the sub-curvature modes but also {\em super-curvature} modes,
with eigenvalues in the range $-1$ to $0$. (An illustration of the need for
super-curvature modes is provided by equation (\ref{opencorr}) below;
it shows that the correlation length, defined as the distance out to which
the correlation function is roughly constant, can be arbitrarily large
if super-curvature modes are included, whereas it is at most of order
the curvature scale if these modes are omitted.)

The Grishchuk--Zel'dovich effect explores by definition scales that
are much bigger than the size of the observable universe, so, ignoring
for the moment the exceptional case that $\Omega_0$ is very close to
one, it necessarily explores very large super-curvature scales
corresponding to an eigenvalue close to zero.  This fact was first
recognized in \cite{LW}, previous analyses \cite{turner,KS,KTF} having
failed to recognize it. On the other hand, in the limit of infinitely
large scales the perturbations become homogeneous and thus can be
represented by superpositions of Bianchi type~V models \cite{Lukash}
which are the anisotropic generalization of the open FRW metric. The
microwave background anisotropy in such models has been discussed
elsewhere \cite{Nature}. The present paper explores the
Grishchuk--Zel'dovich effect in detail for the first time, quantifying
the upper bound that can be placed on the spectrum and explaining its
physical significance.

\section{Mode Functions and Microwave Anisotropies}

The line element for an open universe can be written as
\begin{equation}
ds^2 = -dt^2 + a^2(t)\left[dr^2 + \sinh^2 r \left( d\theta^2 + \sin^2
\theta \, d\phi^2 \right) \right] \,.
\label{line}
\end{equation}
In this expression $a(t)$ is the scale factor of the universe,
normalized so that the spatial curvature scalar is
\begin{equation}
\label{rthree}
R^{(3)}=-6/a^2 \,.
\end{equation}
Space is practically flat on scales much less than $a(t)$ but strongly
curved on much larger scales, so one may call the distance $a(t)$ the
physical {\em curvature scale}. In these units, the comoving curvature
scale is unity.

The Friedmann equation can be written
\begin{equation}
\label{omega}
1-\Omega= \frac 1{(aH)^2} \,,
\end{equation}
where as usual $H=\dot a/a$ is the Hubble parameter, and the
time-dependent quantity $\Omega$ is the energy density measured in
units of the critical density $3H^2/8\pi G$. From this equation it is
clear that the Hubble distance $H^{-1}$ is always less than the
curvature scale.

It is useful to employ conformal time, defined by $\eta \equiv \int
dt/a(t)$, which has the interpretation of being the coordinate
distance to the particle horizon. Its present value $\eta_0$ is an
excellent approximation to the distance to the last scattering surface
and, making the approximation (valid except for very low $\Omega_0$) that the
universe can be treated as matter dominated since last scattering, it is
given by
\begin{equation}
\eta_0 = \cosh^{-1} \left[ \frac{2-\Omega_0}{\Omega_0} \right] \,,
\end{equation}
where subscript `0' always indicates present value. For $\Omega_0 <
2/(1 + \cosh 1) \simeq 0.786$, the surface of last scattering is located
beyond the curvature scale.

In analyzing the Grishchuk--Zel'dovich effect, one needs the full
paraphernalia of mode functions appropriate to an open universe. This
has recently been given in considerable detail by Lyth and Woszczyna
\cite{LW}, and we shall simply repeat the formulae here. The mode
expansion of a generic function $f$ is
\begin{equation}
f(r,\theta,\phi,t)=\int_0^\infty \mbox d k \sum_{lm} f_{klm}(t)
	\Pi_{kl}(r) Y_{lm}(\theta,\phi) \,,
\end{equation}
Here $-(k/a)^2$ is the eigenvalue of the Laplacian corresponding to
the eigenfunction $\Pi_{kl}(r) Y_{lm}(\theta,\phi)$.  For brevity we
shall refer to $k/a$ as the wavenumber, even though one cannot
usefully define plane waves in curved space.

We introduce the notation $q^2=k^2-1$; sub-curvature modes correspond
to $0 < q^2 < \infty$ and super-curvature modes correspond to $-1 <
q^2 < 0$. The angular functions are the usual spherical harmonics. The
radial functions $\Pi_{kl}$ for the sub-curvature modes are given by
\cite{HAR}
\begin{eqnarray}
\Pi_{kl} & \equiv & N_{kl} \tilde \Pi_{kl} \,, \\
\tilde \Pi_{kl} & \equiv & q^{-2} (\sinh   r)^l\left(\frac{-1}{\sinh  r}
	\frac{{\mbox d}}{{\mbox d} r}\right)^{l+1} \cos(q  r) \,, \\
N_{kl} & \equiv & \sqrt\frac2\pi q^2 \prod_{n=0}^l (n^2+q^2)^{-1/2} \,.
\end{eqnarray}
For the super-curvature modes they can be obtained by analytic
continuation
\cite{LW}
\begin{eqnarray}
\Pi_{kl} & \equiv & N_{kl}\tilde\Pi_{kl} \,, \\
\tilde \Pi_{kl} & \equiv & |q|^{-2} (\sinh r)^l\left(\frac{-1}{\sinh  r}
	\frac{{\mbox d}}{{\mbox d} r}\right)^{l+1} \cosh(|q| r) \,, \\
N_{kl} & \equiv & \sqrt\frac2\pi |q| \prod_{n=1}^l (n^2+q^2)^{-1/2}
	\hspace{3em} (l\geq1) \,.
\end{eqnarray}

To construct a Gaussian random field the amplitude $f_{klm}$ of each
mode is given an independent Gaussian probability distribution, whose
variance is defined by the spectrum ${\cal P}_f(k)$ through
\begin{equation}
\label{fullspec}
\langle f^*_{klm} f_{k'l'm'}\rangle = \frac{2\pi^2}{k|q|^2} {\cal P}_f(k)
	\delta(k-k') \delta_{ll'} \delta_{mm'} \,,
\end{equation}
where the angle brackets denote the ensemble average. It determines the
two-point correlation function through the relation
\begin{equation}
\label{opencorr}
\langle f(1)f(2) \rangle = \int^\infty_0 \frac{\mbox d k}{k}
	{\cal P}_f(k) \frac{\sin(q  r)}{q\sinh  r} \,.
\end{equation}
The correlation function depends only on the geodesic distance $ar$
between the comoving points 1 and 2 provided that the spectrum is
independent of $l$ and $m$, and the Gaussian field is then said to be
homogeneous (with respect to the group of transformations leaving the
distance invariant). Evaluated at $r=0$ it gives the
position-independent mean-square
\begin{equation}
\label{mssc}
\langle f^2 \rangle = \int_0^\infty \frac{{\mbox d} k}{k}{\cal P}_f(k) \,.
\end{equation}

We are interested in the perturbation $\delta R^{(3)}_{klm}$ in the
curvature scalar of comoving hypersurfaces (those orthogonal to
comoving worldlines), away from its average value $6/a^2$. It is
conveniently characterized by a quantity ${\cal R}$ defined by
\begin{equation}
4(k^2+3) {\cal R}_{klm}/a^2 =\delta R^{(3)}_{klm} \,.
\label{rdef}
\end{equation}
After matter domination (which is the only era that concerns us)
${\cal R}$ is practically constant until $\Omega$ breaks away from 1.
Before that happens, ${\cal R}$ is equal to $- 5/2$ times the gauge
invariant gravitational potential $\Phi$ \cite{Bardeen},
 but afterwards $\Phi$ is multiplied by a factor
$F(\eta)$ defined by
\begin{equation}
\label{FETA}
F(\eta) = 5\,\frac{\sinh^2\eta-3\eta\sinh\eta+4\cosh\eta-4}
	{(\cosh\eta-1)^3}\,.
\end{equation}
As long as it is constant, ${\cal R}$ is related to the density perturbation
on comoving hypersurfaces by
\begin{equation}
\frac{\delta\rho}{\rho}= \frac{2}{5} \, \frac{k^2+3}{a^2H^2} \,
	{\cal R}_{klm} \,.
\label{denspert}
\end{equation}
{}From now on $\cal R$ will refer exclusively to the constant, early time
value.

The only important effect on the cosmic microwave background anisotropy from
large scales will be gravitational, through the usual Sachs--Wolfe effect.
As the usual practice is to work with the monopole and dipole subtracted from
measured anisotropies, we shall concentrate on the multipoles from the
quadrupole ($l=2$) upwards. However, one must also consider the question of
whether or not the dipole induced by the perturbations is compatible with
observations, and we discuss this in the Appendix.

Following Lyth and Woszczyna \cite{LW}, the appropriate expression (for $l
\geq 2$) is
\begin{equation}
\label{openswsc}
l(l+1)C_l = 2\pi^2 l(l+1) \int^\infty_0 \frac{{\mbox d} k}{k}
	{\cal P}_{\cal R}(k) I_{kl}^2 \ ,
\end{equation}
where $C_l$ is the radiation angular power spectrum (defined as usual
as the ensemble average of the $l$-th multipole of the temperature
anisotropy), and ${\cal P}_{\cal R}$ is the spectrum of the
time-independent primordial curvature perturbation. The function
$I_{kl}^2$ is the `window function' which indicates how a given scale
$k$ contributes to the $C_l$. It is given by
\begin{equation}
\label{window}
q I_{kl} = \frac{1}{5} \,\Pi_{kl}(\eta_0) + \frac{6}{5} \int^{\eta_0}_0
	\mbox d r \,\Pi_{kl}(r)\,F'(\eta_0-r) \,,
\end{equation}
where in calculating the $I_{kl}$, one must use the mode expansion
appropriate to whether the mode is sub-curvature or super-curvature.

In order to uncover the complete form of the window functions for
$\Omega_0 < 1$ it is necessary to evaluate them numerically, both for
sub- and super-curvature modes. We show the window functions
corresponding to the first three multipoles in Figure 1, for the case
$\Omega_0 = 0.2$. As expected, they behave smoothly across the
curvature scale $k = 1$.

The window functions show that if the spectrum continues to be flat or
falling as one goes to larger scales, then there will be little effect
from the large scale modes, and so the large angle cosmic microwave
anisotropy will indeed be dominated by modes of order the Hubble
scale. Only if the spectrum rises can the effect of very large scales
be significant. This is true regardless of the value of $\Omega_0$.

\section{The sub-curvature scale Grishchuk--Zel'dovich effect}

We are interested in the effect on the microwave anisotropies of
scales far bigger than the observable universe. If $\Omega_0$ is close
to one the curvature scale becomes large compared with the size of the
observable universe, giving the possibility that the large scales can
still be in the sub-curvature regime. We consider this first, moving
on to the question of the super-curvature regime in the following
Section. We shall see that in the limit $\Omega_0 \to 1$, only the
sub-curvature effect is important.

As we have chosen the scale factor $a$ to be equal to the curvature
scale, the limit $\Omega_0=1$ corresponds to the limit $a \to \infty$.
Since a physical wavenumber is $k/a$ and a physical radial distance is
$ar$, it also corresponds to $k \to \infty$ and $r \to 0$ with $kr$
fixed.  One can identify $q$ with $k$ in this limit, and the radial
functions are related to the spherical Bessel functions by
\begin{equation}
\label{jlim}
\Pi_{kl}(  r)\to \sqrt\frac{2}{\pi} \, k j_l(k  r) \,.
\end{equation}
The correlation function expression becomes the usual
\begin{equation}
\label{flatcorr}
\langle f(1)f(2) \rangle = \int^\infty_0 \frac{\mbox d k}{k}
	{\cal P}_f(k) \frac{\sin(k  r)}{kr} \,,
\end{equation}
and the relation between ${\cal R}$ and the curvature scalar becomes
\begin{equation}
\label{tweiflat}
4 k^2 {\cal R}_{klm}/a^2 = \delta R^{(3)}_{klm} \,.
\end{equation}

The integral in Eq.~(\ref{window}) vanishes and one finds
\begin{equation}
\label{clflat}
l(l+1) C_l= \frac{4\pi}{25} l(l+1) \int_0^{\infty} \frac{\mbox d k }{k} \,
	{\cal P}_{\cal R}(k)\, j_l^2 (\eta_0 k) \,.
\end{equation}
If the curvature perturbation spectrum is scale independent this gives a
constant value of $l(l+1)C_l$
\begin{equation}
l(l+1)C_l=\frac{2\pi}{25}{\cal P}_{\cal R} \,,
\end{equation}
which is consistent with the COBE data \cite{COBE}.

We represent the effect of large scale modes by a delta function
contribution to the spectrum. We consider a very large scale {\em
sub-curvature} mode
\begin{equation}
\label{vvlarge}
{\cal P}_{{\cal R}}^{\rm SUB} \simeq \delta(\ln k- \ln k_{\rm SUB})
	\langle {\cal R}^2 \rangle _{\rm SUB} \,,
\end{equation}
where
\begin{equation}
1 \ll k_{\rm SUB} \ll a_0H_0= (1-\Omega_0)^{-1/2} \,.
\label{ksub}
\end{equation}
Since $j_l(x) \propto x^l$ as $x \to 0$, the quadrupole dominates all
other multipoles and is given by
\begin{equation}
\label{subquad}
6 C^{\rm SUB}_2 = \frac{148\pi}{1875} \left(\frac{k_{\rm
	SUB}}{a_0H_0}\right)^4 \langle {\cal R}^2 \rangle_{\rm SUB} \,.
\end{equation}
This, in more modern and precise notation, is the result derived by
Grishchuk and Zel'dovich \cite{GZ} for the case $\Omega_0=1$.

In the COBE data the quadrupole is not markedly higher than the others
(quite the reverse if anything), so we take the observational value of
$l(l+1)C_l$ as an upper limit on the effect. This gives
\begin{equation}
\label{climit2}
l(l+1) C_l^{\rm SUB}<8 \times 10^{-10} \,.
\end{equation}
We have taken the observational limit as that corresponding to a flat
spectrum evaluated for $\Omega_0=1$ with an upper limit of $20\mu$K on
the expected quadrupole $Q = \sqrt{5C_2/4\pi}$ \cite{Gorski}. This
leads to the bound
\begin{equation}
\label{omega1con}
\left( \frac{k_{{\rm SUB}}}{a_0H_0}\right)^2 \langle {\cal R}^2
	\rangle^{1/2}_{{\rm SUB}} < 6 \times 10^{-5}\,.
\end{equation}

This bound was derived under the assumption that $k_{{\rm SUB}}$ is a
large {\em sub-curvature} scale, $k_{{\rm SUB}} \gg 1$, which is
consistent with Eq.~(\ref{omega1con}) only if $1-\Omega_0 \lesssim
10^{-4} \langle {\cal R}^2 \rangle^{-1/2}_{{\rm SUB}}$. As we shall
see in Section \ref{phys}, the biggest permissible geometry distortion
corresponds to $\langle {\cal R}^2 \rangle_{{\rm SUB}}\sim 1$, and
with this value consistency requires $1-\Omega_0 \lesssim 10^{-4}$. In
words, a curvature perturbation of order unity on sub-curvature scales
is allowable only if $\Omega_0$ is very close to one \cite{KTF}. This
is to be expected, since otherwise there are no sub-curvature scales
much bigger than the observable universe.

\section{The super-curvature scale Grishchuk--Zel'dovich effect}

If $\Omega_0$ is not close to one, scales far larger than the
observable universe are necessarily much bigger than the curvature
scale, corresponding to super-curvature modes with $0<k\ll 1$. We
again consider a delta function power spectrum, given by
\begin{equation}
\label{dfps}
{\cal P}_{\cal R}^{\rm VL} \simeq \delta(\ln k-\ln k_{\rm VL})
	\langle {\cal R}^2 \rangle_{\rm VL} \,,
\end{equation}
where $k_{\rm VL}$ is a scale satisfying $0<k_{\rm VL} \ll 1$. The
effect of such a contribution was investigated qualitatively in
Refs.~\cite{LW,Munich}, but here we present a full quantitative
calculation and also explain more fully the physical significance of
the result.

The limit of small $k$ can be taken partly analytically, though one
can obtain the same results by numerical calculation from the full
expressions given above. In this limit, $\Pi_{k0} \to 1$, but the
other radial functions are proportional to $k$. It is convenient to
define
\begin{eqnarray}
N_l &\equiv& \lim_{k\to0}\ k\,N_{kl} \ ,\\
\tilde\Pi_l &\equiv& \lim_{k\to0}\ \tilde\Pi_{kl}/k^2 \ ,
\end{eqnarray}
from which one finds
\begin{equation}
N_l = \sqrt{\frac2\pi} \,\prod_{n=2}^l(n^2-1)^{-1/2}
	\hspace{5mm}(l\geq 2) \ .
\end{equation}
and
\begin{eqnarray}
\label{PI12}
\tilde\Pi_{1}(r) &=& {1\over2}\left[\coth\,r - {r\over\sinh^2 r}\right]
	\,, \\
\tilde\Pi_{2}(r) &=& {1\over2}\left[1 +
	{3(1 - r\,\coth\,r)\over\sinh^2 r}\right] \,.
\end{eqnarray}
The other radial functions follow from the recurrence relation
\begin{equation}
\tilde\Pi_{l}(r) = - l(l-2)\,\tilde\Pi_{l-2}(r) + (2l-1)\,\coth\,r\
	\tilde\Pi_{l-1}(r)\,.
\end{equation}
Using these results, the contribution to the mean square multipoles
becomes
\begin{equation}
\label{clvl}
l(l+1)\,C_l^{\rm VL} = l(l+1)\,N_l^2\,B_l^2\, k^2_{\rm VL}
	\langle{\cal R}^2\rangle_{\rm VL} \,,
\end{equation}
where
\begin{equation}
B_{l} \equiv \frac{1}{5} \, \tilde\Pi_l(\eta_0)+\frac{6}{5} \,
	\int^{\eta_0}_0 {\rm d}r\,\tilde\Pi_l(r)\,F'(\eta_0-  r)\ .
\end{equation}
By evaluating the full expressions in Eq.~(\ref{openswsc})
numerically, we have found that these limits are an excellent
approximation for calculating $C_l$, at least for the low multipoles
we consider. For $k_{\rm VL} \leq 0.1$ the approximation is good to
within a few percent for any reasonable $\Omega_0$.

\subsection{The shape of the spectrum}

In the case $\Omega_0=1$ the Grishchuk--Zel'dovich effect comes
entirely from the sub-curvature modes that we discussed in the
previous Section, and is present only in the quadrupole. It was noted
in Refs.~\cite{LW,Munich} that for $\Omega_0<1$, on the other hand,
the super-curvature Grishchuk--Zel'dovich effect is present in all
multipoles\footnote{As $\Omega_0$ becomes small, the actual pattern produced
by a single infinite scale mode becomes a hot (or cold) spot on the sky of
angular size $\sim 80\, \Omega_0$ degrees \cite{Nature}.}. However, in those
papers the $l$ dependence was not evaluated.

For $\Omega_0$ close to one, it can be shown analytically that $C_l^{\rm VL}
\propto (1-\Omega_0)^l$, so that the quadrupole dominates.\footnote{This
result that the quadrupole dominates refers to the very large scale
super-curvature modes corresponding to $k\ll 1$. In the previous Section we
found that the quadrupole also dominates for large scale sub-curvature modes
corresponding to $1\ll k\ll a_0H_0$. It also dominates for the intermediate
scales $k \sim 1$; the origin of this result is that for small $r$, both sub-
and super-curvature modes have the same leading behaviour when expanded in
small $|q|r$.} The prefactor can be evaluated analytically, yielding
\begin{eqnarray}
\label{C2KR}
6 C_2^{\rm VL} & \simeq & \frac{64}{625\pi} \, k^2_{{\rm VL}}
	\langle{\cal R}^2\rangle_{\rm VL} \, (1-\Omega_0)^2 \,, \\
\label{C3KR}
12 C_3^{\rm VL} & \simeq & \frac{4096}{30625\pi} \, k^2_{{\rm VL}}
	\langle{\cal R}^2\rangle_{\rm VL} \, (1-\Omega_0)^3 \,.
\end{eqnarray}
We have confirmed that our numerical code reproduces these
coefficients. It demonstrates that these expressions are very accurate
for $\Omega_0 > 0.9$, and still a reasonable approximation for
$\Omega_0 = 0.8\,$.

For $\Omega_0$ significantly below one, numerical calculation is
essential. In Figure 2 we show the shape of the radiation power
spectra for several different values of $\Omega_0$, normalized such
that all have the same $C_2$.  One sees a complicated behaviour, which
is not so surprising since standard calculations of the multipoles
using spectra of sub-curvature modes already show a very complex
behaviour with $\Omega_0$ even if only the Sachs-Wolfe terms are
included \cite{KS,KRSS}.  Unusual behaviour is particularly marked for
$\Omega_0 \simeq 0.35$; an `accidental' suppression of the quadrupole
even from curvature-scale perturbations, due to cancellation between
the `intrinsic' and `line-of-sight' terms, has already been noted
\cite{KTF}. For this $\Omega_0$ the contribution of very large scales
to the octopole is considerably larger than that to the quadrupole.

A question one would like to address is whether for any $\Omega_0$ the
shape induced by the super-curvature modes resembles the shape observed by
COBE. A convenient way of characterizing the observed shape is to note that
it is compatible with the one induced in an $\Omega_0=1$ universe by
power-law spectra of density perturbations, with the spectral index $n$
roughly constrained to a range 0.7 to 1.4 about the Harrison--Zel'dovich
value $n=1$ \cite{Gorski}. As we have remarked, the assumption of $n=1$ leads
to $l(l+1) C_l$ being constant. We have illustrated the allowed slopes of
the spectra in Figure 2 for comparison.

We find that over very short ranges of multipoles it is possible for
the Grishchuk--Zel'dovich effect to mimic the observations, which is
not possible in a flat universe. However, for no value of $\Omega_0$
is the spectrum sustained as approximately flat even over just the
limited range up to $l = 10$. We conclude then that the
Grishchuk--Zel'dovich effect cannot be responsible for the entire
shape of the radiation power-spectrum (a conclusion which will
surprise nobody!), but that it remains possible for some fraction of
the measured multipoles to be due to this effect. In the flat space
limit only the quadrupole can be partly generated this way.

\subsection{The amplitude of the spectrum}

Earlier papers \cite{LW,Munich} assumed that the quantities $N_l$ and
$B_l$ are roughly of order one for low multipoles. However, we find
considerable cancellations in $B_l$ and in fact their typical sizes
are somewhat smaller.  Continuing with the choice of a delta function
power spectrum, we evaluate them numerically and compare the
anisotropies to the observational limit, demanding again that they be
less than the observed value
\begin{equation}
l(l+1) C_l^{\rm VL} <8 \times 10^{-10} \,.
\label{climit}
\end{equation}

In Figure 3 we plot the dependence of the $l = 2$, $3$, $4$ and $5$
multipoles on $\Omega_0$. In accordance with Eq.~(\ref{clvl}),
$l(l+1)C_l$ scales with $k^2_{\rm VL} \langle{\cal R}^2\rangle_{\rm
VL}$, so we have plotted it normalized by this quantity. Near
$\Omega_0 = 1$, the analytic approximations of Eqs.~(\ref{C2KR}) and
(\ref{C3KR}) can be used.

By taking the upper limits on the Grishchuk--Zel'dovich effect given
above, one can obtain a maximum permitted value for $k^2_{\rm VL}
\langle{\cal R}^2\rangle_{{\rm VL}}$, as a function of $\Omega_0$, and
this is plotted in Figure 4. We have imposed the constraint on
multipoles up to $l = 6$ rather than just the quadrupole, since the
`accidental' cancellation of the quadrupole for $\Omega_0 \simeq
0.35$ would otherwise distort the constraint.

We see from Figure 4 that for $0.25 \lesssim \Omega_0 \lesssim 0.8$
the bound is fairly constant
\begin{equation}
k^2_{\rm VL}\langle{\cal R}^2\rangle_{\rm VL} \lesssim 10^{-6} \,.
\label{omegasmall}
\end{equation}
For $\Omega_0<0.25$  it becomes more severe. On the other
hand, for $0.8 \lesssim \Omega_0 < 1$, the dependence of Eq.~(\ref{C2KR})
yields
\begin{equation}
k_{{\rm VL}}^2 \langle{\cal R}^2\rangle_{{\rm VL}} < \frac{3 \times
        10^{-8}} {(1-\Omega_0)^2} \,.
\label{omegabig}
\end{equation}
These bounds were derived under the assumption that $k_{{\rm VL}} \ll
1$, so Eq.~(\ref{omegasmall}) is automatically satisfied if
$\langle{\cal R}^2\rangle_{{\rm VL}} \lesssim 10^{-6}$ and
Eq.~(\ref{omegabig}) is automatically satisfied if $\langle {\cal R}^2
\rangle_{{\rm VL}} \lesssim 10^{-8}/(1-\Omega_0)^2$. As we shall see
later, the biggest permissible geometry distortion corresponds to
$\langle{\cal R}^2\rangle_{{\rm VL}} \sim 1$. Even with this maximal
value, Eq.~(\ref{omegabig}) is automatically satisfied if $1-\Omega_0
\lesssim 10^{-4}$; in this regime, the maximal geometry distortion is
allowed on all super-curvature scales with $k_{{\rm VL}} \ll 1$. This
result is hardly surprising, since super-curvature scales move off to
infinity in the limit $\Omega_0 \to 1$.

The bound can also be written in terms of $k/(aH)$, which is the
physical wavenumber in Hubble units. For $0.25 \lesssim \Omega_0
\lesssim 0.8$ it becomes
\begin{equation}
\label{omegasmall2}
\frac{k^2_{\rm VL}}{a_0^2 H_0^2} \langle{\cal R}^2\rangle_{\rm VL}
	\lesssim 10^{-6} (1-\Omega_0) \,,
\end{equation}
and for $0.8 \lesssim \Omega_0 < 1$ it becomes
\begin{equation}
\label{omegabig2}
\frac{k^2_{\rm VL}}{a_0^2 H_0^2} \langle{\cal R}^2\rangle_{{\rm VL}}
	\lesssim \frac{3 \times 10^{-8}}{1-\Omega_0} \,.
\end{equation}
Provided $\Omega_0$ is not too close to one, then, assuming the
maximal value $\langle{\cal R}^2\rangle_{{\rm VL}} \sim 1$, these
results say that that the inverse wavenumber must be at least about a
thousand times the Hubble distance (for $\Omega_0 > 0.25$). As
$\Omega_0$ approaches one, the bound weakens slowly. At $1-\Omega_0
\simeq 10^{-4}$, it requires the inverse wavenumber to be at least a
factor of about a hundred times the Hubble distance, and by this time
super-curvature scales are far enough outside the observable universe
to automatically satisfy this requirement.

\section{Physical interpretation of the bounds}
\label{phys}

In order to interpret these bounds physically, we have to explain the
meaning of the curvature perturbation ${\cal R}$, showing that within
the context of a homogeneous Gaussian random field it can be at most
of order one.

For this physical interpretation, we assume that the observed
curvature perturbation is a typical realization of a homogeneous
Gaussian random field not only within the observable universe, but
also in a far bigger region, much bigger in fact than the correlation
length.

Even though the background space has negative curvature, the presence
of a large perturbation can render regions of the universe positively
curved.  Moreover, within a region no bigger than the correlation
length the curvature is practically homogeneous so that we are dealing
with a practically homogeneous space just as in the unperturbed
universe.  But in a homogeneous space with positive curvature scalar
$R^{(3)}$, the biggest sphere that can be drawn has radius $d$ given
by\footnote{This equation follows from the positive-curvature version
of Eq.~(\ref{line}), given by $\sinh\to \sin$, corresponding to the
maximum value $r=\pi$.  When it is small, $R^{(3)} d^2$ is a linear
measure of the geometry distortion within the sphere so that for
instance the area of the sphere is equal to $(1-2R^{(3)} d^2)$ times
its Euclidean value $4\pi d^2$.}
\begin{equation}
R^{(3)} d^2=\pi^2/6\sim 1 \,.
\end{equation}
In other words, positive curvature inside a given sphere cannot exceed
this value. One can say that for bigger positive curvature, space
would `close in on itself', indicating that our underlying assumptions
break down. What we are going to do is demonstrate that, at least for
a perturbation coming solely from a single scale, this requirement is
equivalent to ${\cal R}\lesssim 1$.

The situation is
different for sub- and super-curvature scales, and we begin by
considering the former. In that case the correlation length $d_{\rm
VL}$, defined as the distance $ar$ within which correlation function
Eq.~(\ref{flatcorr}) is practically constant, is simply the inverse
wavenumber
\begin{equation}
d_{\rm SUB}=a\, k_{\rm SUB}^{-1} \,.
\end{equation}
Within a typical sphere of this radius ${\cal R}$ is practically constant,
and the curvature scalar given by Eq.~(\ref{rdef}) is
\begin{equation}
4 {\cal R}= d_{\rm SUB}^2 \delta R^{(3)}\, .
\end{equation}
If ${\cal R}\sim 1$ the perturbation $\delta R^{(3)}$ always dominates
the background value $R^{(3)}=-6/a^2$ because $k_{\rm SUB}\gg 1$.

According to the linear cosmological perturbation theory that we are
invoking in this paper, the actual value of $\cal R$ within a randomly
located sphere of fixed radius has a Gaussian probability
distribution.\footnote{In referring to a `randomly located' sphere we are
taking for granted the ergodic property, which is essentially the statement
that choosing a randomly located sphere for a fixed realization of the random
field is equivalent to choosing a random realization of the field at a fixed
location. Only in the latter case is the Gaussian property an immediate
consequence of our assumptions. However the ergodic property can be proved
\cite{adler} under weak assumptions for a Euclidean geometry corresponding to
$\Omega_0=1$, and we see no reason to suppose that it fails for the case of
curved space.} Taken literally the linear theory therefore predicts the
existence of regions in which ${\cal R}\gg 1$, in contradiction with the
physical interpretation of that quantity. It follows that at least within the
framework that we are adopting the mean square contribution $\langle {\cal R}
\rangle^2_{\rm SUB}$ cannot exceed one.\footnote{In what follows we assume
that linear cosmological perturbation theory is valid right up to this
maximum value, or in other words that it is valid as long as it predicts only
rare regions of space with the unphysical value ${\cal R} \gtrsim 1$. This
seems very likely because there is no obvious criterion which would indicate
earlier non-linearity. In particular, on the large scales that we are
considering the fractional density perturbation Eq.~(\ref{denspert})
on comoving hypersurfaces is small if ${\cal R}$ is small.} With this
maximal value, Eq.~(\ref{omega1con}) becomes
\begin{equation}
d_{\rm SUB} \gtrsim 130 \, H_0^{-1} \,.
\end{equation}

This discussion of the sub-curvature case has led to no surprises. Consider
now the super-curvature case, corresponding to the scale $k_{\rm VL}\ll 1$.
{}From Eq.~(\ref{opencorr}) it follows that the correlation length is now
given by
\begin{equation}
d_{\rm VL}=a_0 \, k_{\rm VL}^{-2} \,.
\end{equation}
This differs from the Euclidean result through the appearance of $k^2$
instead of $k$. As pointed out in \cite{LW}, the different power can be
understood from the different relation between volume and area.\footnote{In
Eq.~(50) of Ref.~\cite{LW}, the symbol $f$ should be replaced by $\xi$.}

Within a sphere whose radius is of order $d_{\rm VL}$, the correlation
length ${\cal R}$ is practically constant and so is the corresponding
curvature scalar given by Eq.~(\ref{rdef}) as
\begin{equation}
12 {\cal R} \simeq a^2 \delta R^{(3)} \,.
\end{equation}
Since the background curvature is $R^{(3)}=-6/a^2$ we see that $2{\cal
R}$ is equal to the {\em fractional} change in the geometry
distortion. It will be negative in some regions and positive in
others, but on scales much bigger than the curvature scale the
background distortion is huge (the ratio of area to volume is
exponentially large compared with the Euclidean ratio). As a result
the maximum allowed positive value of ${\cal R}$ is again of order one
and we again require for the mean square $\langle {\cal R}^2
\rangle_{{\rm VL}} \lesssim 1$. But this maximal value now corresponds
to a huge distortion of the geometry.

For the maximal super-curvature geometry distortion $\langle{\cal R}
\rangle^2_{\rm VL} \sim 1$, the correlation length must satisfy
\begin{equation}
d_{\rm VL}\gtrsim 10^6 \, H_0^{-1} (1-\Omega_0)^{-1/2} \,,
\end{equation}
if $0.25 \lesssim \Omega_0 \lesssim 0.8$, and
\begin{equation}
\label{supcons}
d_{\rm VL} \gtrsim 4\times 10^7 \, H_0^{-1}(1-\Omega_0)^{3/2} \,,
\end{equation}
if $\Omega_0 \gtrsim 0.8$. As stated earlier, this second bound is
automatically satisfied in the extreme case $1-\Omega_0\lesssim 10^{-4}$.

To summarize, if there is a maximal geometry distortion on a
super-curvature scale, then this scale must be more than a million
times the Hubble distance if $\Omega_0=0.25$. As $\Omega_0$ rises the
constraint weakens, but only in the extreme regime $1-\Omega_0\lesssim
10^{-4}$ does it disappear completely, so that all super-curvature
correlation lengths $d_{\rm VL}$ are allowed.  This is just the regime
in which one can have maximal geometry distortion also on
sub-curvature scales, provided that the correlation length is more
than a hundred times the Hubble distance.

\section{Discussion}

While it remains possible that some component of the observed cosmic
microwave background anisotropies is due to the Grishchuk--Zel'dovich
effect, there is no evidence that this is the case and so at present
one is left with a null result. In absolute generality, this null
result tells us nothing at all about the nature of the universe beyond
our horizon. This is because the observed anisotropies are due
entirely to the spatial and temporal variation of the curvature
perturbation at or within the last scattering surface; anything could
happen to it immediately outside our observable universe without any
effect being noticeable. In order to say anything further, one must
adopt the hypothesis that even far beyond the observable universe the
curvature perturbation is a typical realization of a homogeneous
Gaussian random field. In that case one expects to be able to break
the perturbation up into modes {\em including} very large scale modes,
and the Grishchuk--Zel'dovich effect then dictates how rapidly the
perturbations may reach large amplitudes as one goes to large scales.
Although there is no way of testing whether the random field
hypothesis is actually correct, the idea that the curvature
perturbation is of such a form is the basis for most work in large
scale structure. It makes particular sense in the case of a flat
universe; remembering that the horizon scale at last scattering is
much smaller than at present, it would be strange for the hypothesis
to only break down at precisely the present epoch. In the case of an
open universe, the hypothesis seems {\em a priori} more restrictive,
because the geometry has selected out a special scale, the curvature
scale, beyond which different physics might apply.

While our results have been framed in a particularly general manner,
it is interesting to consider their implications in the context of
open universe inflationary scenarios. The motivation in the early
papers on the open universe Grishchuk--Zel'dovich effect
\cite{turner,KS,KTF} was an attempt to render unlikely the prospect of
open universe inflation, which at that time was modeled as chaotic
inflation of sufficiently short duration that the universe was not
forced to spatial flatness \cite{ELM}. This was on the grounds that
such inflation could not explain the homogeneity of the universe in a
comoving region bigger than the Hubble distance at the beginning of
inflation (and hence on any comoving scale where the curvature is
significant). However, in the strictest sense as discussed above, the
absence of the Grishchuk--Zel'dovich effect cannot be construed as
evidence that the universe is indeed homogeneous in such a
region. These papers also failed to provide a complete treatment;
Turner \cite{turner} ignored spatial curvature and none of
Refs.~\cite{turner,KS,KTF} included super-curvature modes.

More interesting are the recently discussed `single-bubble' models of open
inflation \cite{hist,STYY,STY,BGT,LM,YST}, which are capable of erasing large
scale perturbations even above the curvature scale while still resulting in
an open universe. This undermines the original motivation of the early
work on the Grishchuk--Zel'dovich effect, but at the same time offers
a new motivation, because it appears possible for super-curvature
modes to be generated in such models.  Indeed, in Ref.~\cite{STY} a
primordial spectrum for such models is calculated and, provided the
inflaton mass is sufficiently small, it includes a single discrete
super-curvature mode (which approaches $k = 0$ in the massless limit) as well
as a continuum of sub-curvature modes. This ties in nicely with our
delta-function analysis. In Ref.~\cite{YST} the microwave anisotropies for
such a composite spectrum are calculated, and as far as we can ascertain from
their figures their results are in good agreement with ours.

In conclusion then, we have confirmed that, even in an open universe,
the low multipoles of the observed microwave anisotropies cannot be due
solely to an adiabatic density perturbation on scales much larger than the
observable universe, given the usual assumption that the perturbation is a
typical realization of a homogeneous Gaussian random field. We have also
provided limits as to how quickly the typical amplitude of perturbations can
reach unity as one moves to larger scales. Although the entire observed
anisotropies cannot be due to the Grishchuk--Zel'dovich effect, it remains
possible that some component of them does have such an origin. This may be
the case in the `single-bubble' models of open inflation.

\section*{Acknowledgements}
JGB and DW are supported by PPARC (UK) and ARL by the Royal Society. We thank
John Barrow and Leonid Grishchuk for discussions.

\appendix
\section*{The dipole contribution}

In this paper so far, we have made the usual assumption that the monopole and
dipole are subtracted from observations before a comparison with
theory is made. The motivation for ignoring the dipole is that in standard
flat space scenarios with scale-invariant density perturbations, the observed
microwave anisotropy dipole will be dominated by our own peculiar motion
relative to the last scattering surface as induced by nearby perturbations,
and hence the `Sachs--Wolfe' microwave dipole is not observable
\cite{brly}. However, a large scale perturbation such as the one we are
considering will generate both Sachs--Wolfe and peculiar velocity dipoles,
and one should check that these are indeed small as compared to the observed
dipole, a velocity of about $400 \, {\rm km} \, {\rm s}^{-1}$ \cite{COBE}
which is equivalent to $C_1 \simeq 2 \times 10^{-6}$, in order to be sure
that ignoring the dipole is justified.

Considering only the Sachs--Wolfe dipole, given by the $l=1$ version of
Eq.~(\ref{openswsc}), it seems on the face of it that the worst problem is in
the flat space limit, since $C_1 \propto (1-\Omega_0)$, while the higher
$C_l$ are proportional to $(1-\Omega_0)^l$ [see Eqs.~(\ref{C2KR}) and
(\ref{C3KR})]. It therefore appears that the dipole will be much larger than
the quadrupole and thus may provide a stronger constraint. However, it is
well known that in flat space the Sachs--Wolfe dipole from a very large
perturbation is exactly cancelled by the `peculiar velocity' dipole caused by
our peculiar motion induced by the same perturbation \cite{brly}. For
$\Omega_0 \leq 1$, our present (physical) peculiar velocity is given by
\cite{MFB}
\begin{equation}
v_i(\eta_0) = - \frac{2}{3} \, G(\eta_0) \, {\cal R},_{i} \,,
\end{equation}
where
\begin{equation}
G(\eta) = \left(\cosh \eta -1 \right) \left[ {2\over5}\,\coth\eta\,F(\eta)
	+ {2\over\sinh\eta} \right] \,,
\end{equation}
and $F(\eta)$ is given by Eq.~(\ref{FETA}).
Physically, the picture is one of a gravitational potential gradient across
our observed universe; this leads to a redshift and blueshift of light from
opposite sides of the sky as it propagates towards us, but the same
gravitational potential is responsible for our local peculiar velocity which
for $\Omega_0 = 1$ has an equal and opposite effect. Consequently, the
leading $(1-\Omega_0)$ dependence in the dipole cancels out, and since there
is in fact no $(1-\Omega_0)^2$ term the dipole is of order $(1-\Omega_0)^3$
and is much smaller than the quadrupole as $\Omega_0$ approaches one. It
therefore does not contribute any additional constraint. We have confirmed
this argument by full numerical calculation of the different dipole terms.
Note that since both the sub- and super-curvature modes have the same leading
behaviour in the expansion in $|q|r$, one need not distinguish which regime
we are in.

We now check that the dipole does not give an additional constraint away from
the flat space limit. Here it is easiest to show that both the Sachs--Wolfe
dipole and peculiar velocity dipole are separately small, as we do not need
to rely on a cancellation.

First we consider the Sachs--Wolfe dipole, given by Eq.~(\ref{openswsc}). As
we have already seen, for $\Omega_0$ away from one all the multipoles $l \geq
2$ are of the same order of magnitude, and one expects this to be true also
of the dipole. We have verified this by direct evaluation; the dependence of
the Sachs--Wolfe dipole on $\Omega_0$ is similar to the higher multipoles
which were shown in Figure 3, and it is never much larger than the higher
multipoles.

We have evaluated the peculiar velocity for the same choice of a delta
function power spectrum, and find that if one imposes the constraint from the
quadrupole, e.~g.~Eq.~(\ref{omegasmall}), then the peculiar velocity dipole
is
indeed much smaller than the observed dipole. This is true even for small
$\Omega_0$.

In conclusion then, the anisotropy dipole is not constraining for any
value of $\Omega_0$.



\vspace{.2cm}
\begin{center}
{\bf FIGURE CAPTIONS}
\end{center}

\noindent
{\bf Figure 1:} The window functions $l(l+1)I^2_{kl}$ for $l = 2$, $3$ and
$4$ (from left to right), for the choice $\Omega_0 = 0.2$. The
curvature scale $k = 1$ is indicated by the thin vertical line.

\vspace*{24pt}
\noindent
{\bf Figure 2:} The shape of the radiation angular power spectra $l(l+1)C_l$
induced by a large scale delta function power spectrum, normalized to equal
values of $C_2$. Shown are $\Omega_0 = 0.8$ (stars), $0.5$ (squares), $0.4$
(triangles) and $0.3$ (circles). The solid lines indicate the spectra
obtained in a flat universe from  power-law perturbation spectra with
spectral indices $n =
0.7$ and $1.4$, which roughly indicate the range of slopes favoured by the
COBE data.

\vspace*{24pt}
\noindent
{\bf Figure 3:} The behaviour of the multipoles as a function of $\Omega_0$,
normalized to the scale and magnitude $k_{\rm VL}^2 \langle {\cal R}^2
\rangle$ of the delta function power spectrum. The multipoles $l = 2,3,4,5$
are indicated by solid, dashed, dot-dashed and dotted lines respectively. For
each multipole, there is an $\Omega_0$ for which the `intrinsic' and
`line-of-sight' Sachs-Wolfe terms cancel exactly, which explains the complex
behaviour of the spectral shape shown in Figure 2. The limiting behaviour as
$\Omega_0 \to 1$ can be computed analytically and is given in
Eqs.~(\ref{C2KR}) and~(\ref{C3KR}) for $l=2$ and~$3$.

\vspace*{24pt}
\noindent
{\bf Figure 4:} Limits on $k_{\rm VL}^2 \langle {\cal R}^2 \rangle$,
derived via Figure 3, based on current observational limits on
$l(l+1)C_l<8\times10^{-10}$. This indicates how large a scale must be
before a perturbation of, for example, order unity would not have made
itself apparent in the observed microwave anisotropies. At large and
small $\Omega_0$ the quadrupole provides the strongest constraint; the
gradient discontinuities are where there is a change in the multipole
giving the strongest constraint.


\begin{references}
\bibitem{GZ} L. P. Grishchuk and Ya. B. Zel'dovich, Astron. Zh. {\bf 55},
	209 (1978) [Sov. Astron. {\bf 22}, 125 (1978)].
\bibitem{COBE} G. F. Smoot et al., Astrophys. J. {\bf 396}, L1 (1992); C. L.
	Bennett et al., Astrophys. J. {\bf 436}, 423 (1994).
\bibitem{LW} D. H. Lyth and A. Woszczyna, ``Large scale perturbations in
	the open universe'', to appear, Phys. Rev. D, astro-ph/9501044.
\bibitem{krein} M. G. Krein, Ukrain. Mat. Z. {\bf 1}, No. 1, 64 (1949);
	{\em ibid} {\bf 2}, No. 1, 10 (1950).
\bibitem{turner} M. S. Turner, Phys Rev D{\bf 44}, 3737 (1991).
\bibitem{KS} M. Kamionkowski and D. N. Spergel, Astrophys. J. {\bf 432},
	7 (1994).
\bibitem{KTF} A. Kashlinsky, I. I. Tkachev and J. Frieman, Phys. Rev. Lett.,
	{\bf 73}, 1582 (1994).
\bibitem{Lukash} G. S. Bisnovatyi-Kogan, V. N. Lukash and I. D. Novikov in
	Proc. of the 5th Regional Meet. Astronomy, Liege (IAU/EPS, 1980).
\bibitem{Nature} J. D. Barrow, R. Juszkiewicz and D. H. Sonoda, Nature
	{\bf 305}, 397 (1983); Nature {\bf 316}, 48 (1985); V. N. Lukash and
	I. D. Novikov, Nature {\bf 316}, 46 (1985).
\bibitem{HAR} E. R. Harrison, Rev. Mod. Phys. {\bf 39}, 862 (1967).
\bibitem{Bardeen} J. M. Bardeen, Phys. Rev. D{\bf 22}, 1882 (1980).
\bibitem{Munich} D. H. Lyth, ``The Grishchuk--Zel'dovich Effect in the
	Open Universe'', to appear, proceedings of the 17th Texas symposium
	(1995),	astro-ph/9501113.
\bibitem{KRSS} M. Kamionkowski, B. Ratra, D. N. Spergel and N. Sugiyama,
	Astrophys. J. Lett. {\bf 434}, L1  (1994).
\bibitem{Gorski} K. M. G\'orski et al., Astrophys. J. {\bf 430}, L89 (1994).
\bibitem{adler} R. J. Adler, {\em The Geometry of Random Fields}, (Wiley,
	Chichester, 1981).
\bibitem{ELM} G. F. R. Ellis, D. H. Lyth and M. B. Mijic, Phys. Lett.
	{\bf B271}, 52 (1991); D. H. Lyth and E. D. Stewart, Phys Lett.
	{\bf B252}, 336 (1990).
\bibitem{hist} S. Coleman and F. De Luccia, Phys. Rev. D{\bf 21}, 3305
	(1980); J. R. Gott, Nature {\bf 295}, 304 (1982); J. R. Gott and T.
	S. Statler, Phys. Lett. {\bf B136}, 157 (1984).
\bibitem{STYY} M. Sasaki, T. Tanaka, K. Yamamoto and J. Yokoyama,
	Phys. Lett. {\bf B317}, 510 (1993); T. Tanaka and M. Sasaki, Phys.
	Rev. D{\bf 50}, 6444 (1994).
\bibitem{STY} M. Sasaki, T. Tanaka and K. Yamamoto, Phys. Rev. D{\bf 51},
	2979 (1995).
\bibitem{BGT} M. Bucher, A. S. Goldhaber and N. Turok, ``An Open Universe
	from Inflation'', to appear, Phys. Rev. D, hep-ph/9411206; M.
	Bucher and N. Turok, ``Open Inflation with an Arbitrary False
	Vacuum Mass'', Princeton preprint (1995), hep-ph/9503393.
\bibitem{LM} A. D. Linde, Phys. Lett. {\bf B351}, 99 (1995); A. D. Linde and
	A. Mezhlumian, ``Inflation with $\Omega \neq 1$'', Stanford preprint
	(1995), astro-ph/9506017.
\bibitem{YST} K. Yamamoto, M. Sasaki and T. Tanaka, ``Large Angle
	CMB Anisotropies in an Open Universe in the One-Bubble
	Inflationary Scenario'', Kyoto preprint (1995), astro-ph/9501109.
\bibitem{brly} M. Bruni and D. H. Lyth, Phys. Lett. {\bf B323} 118 (1994).
\bibitem{MFB} V. F. Mukhanov, H. A. Feldman and R. H. Brandenberger,
        Phys. Rep. {\bf 215}, 203 (1992).
\end{references}
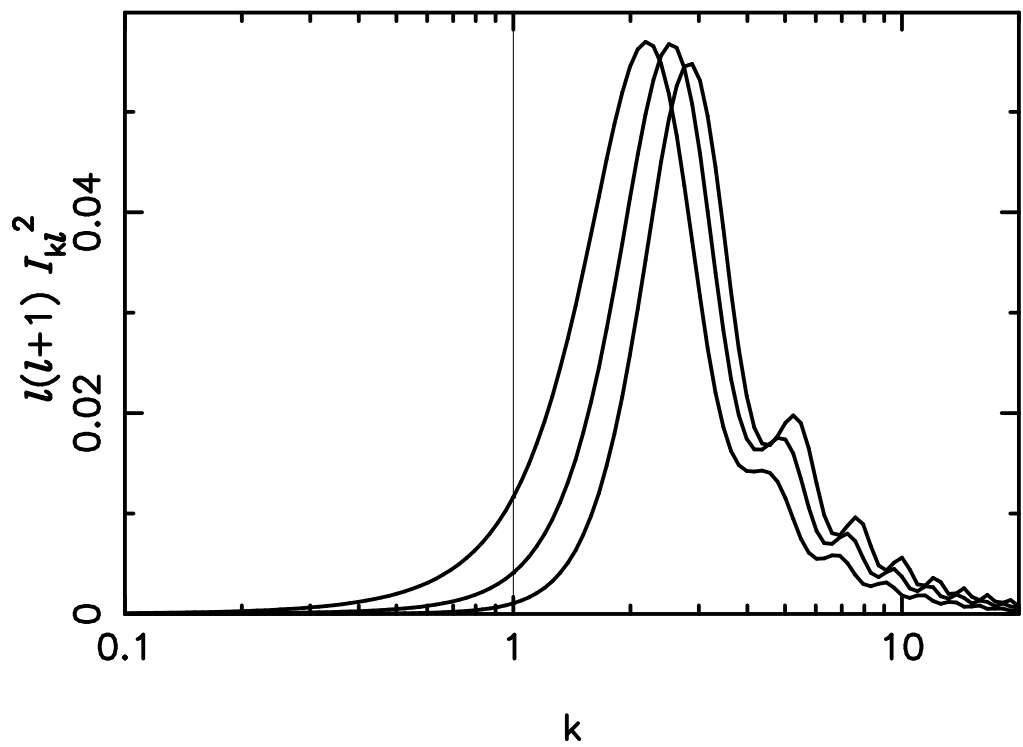
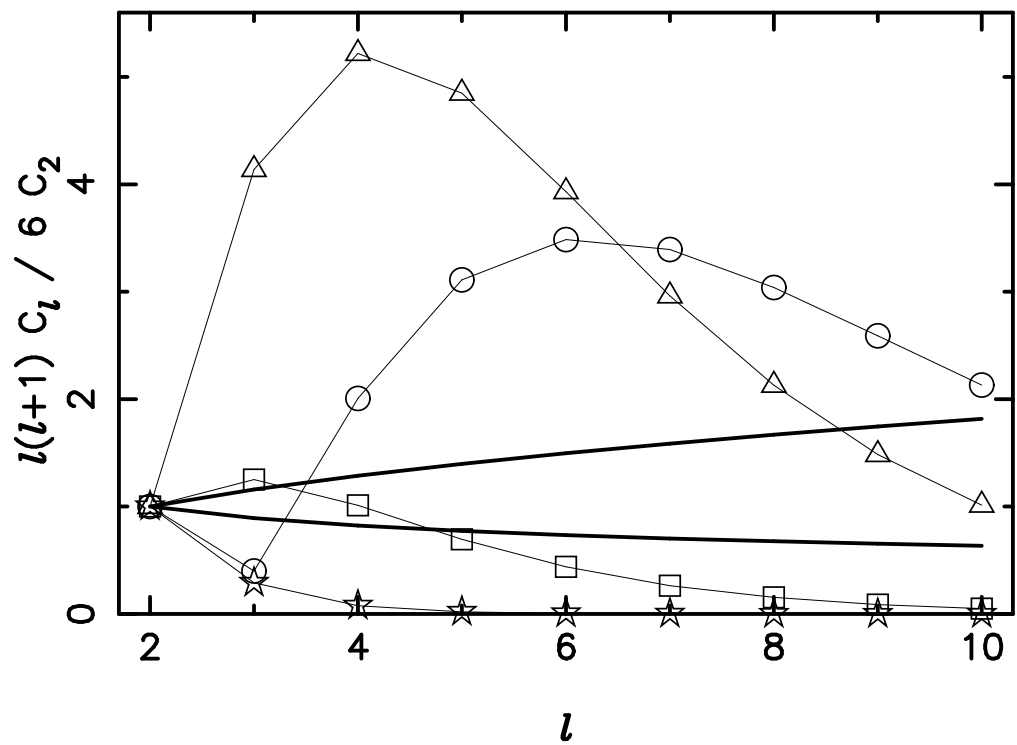
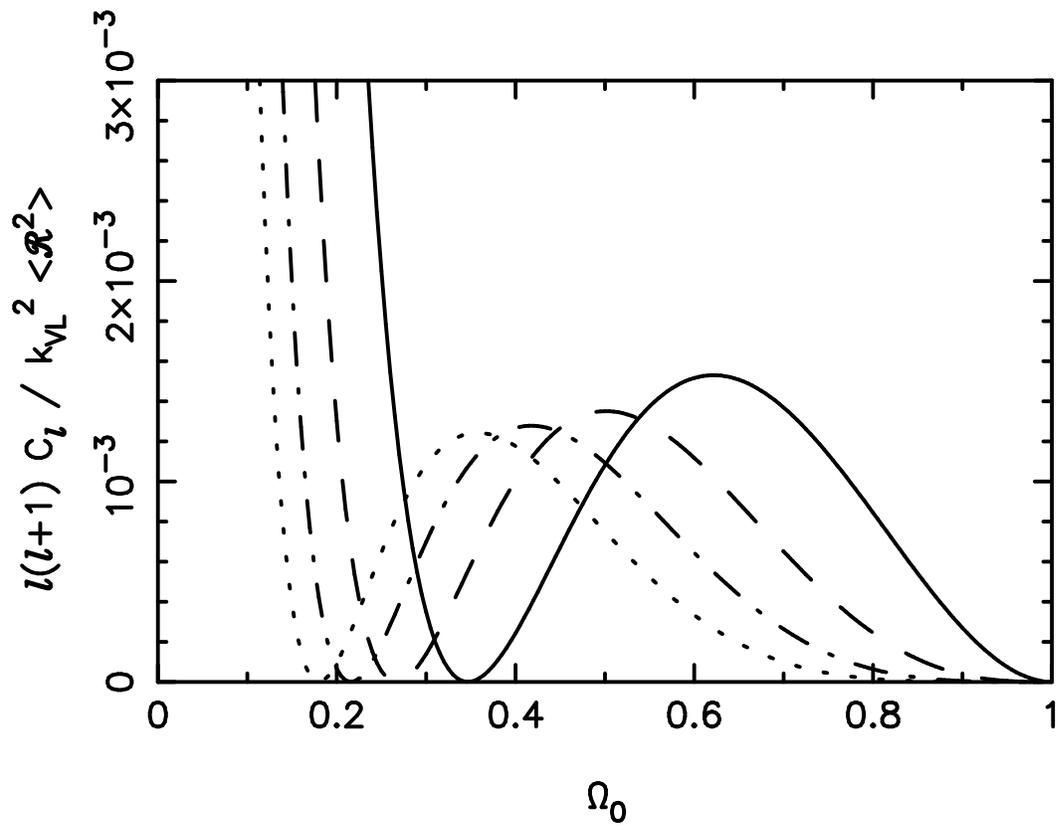
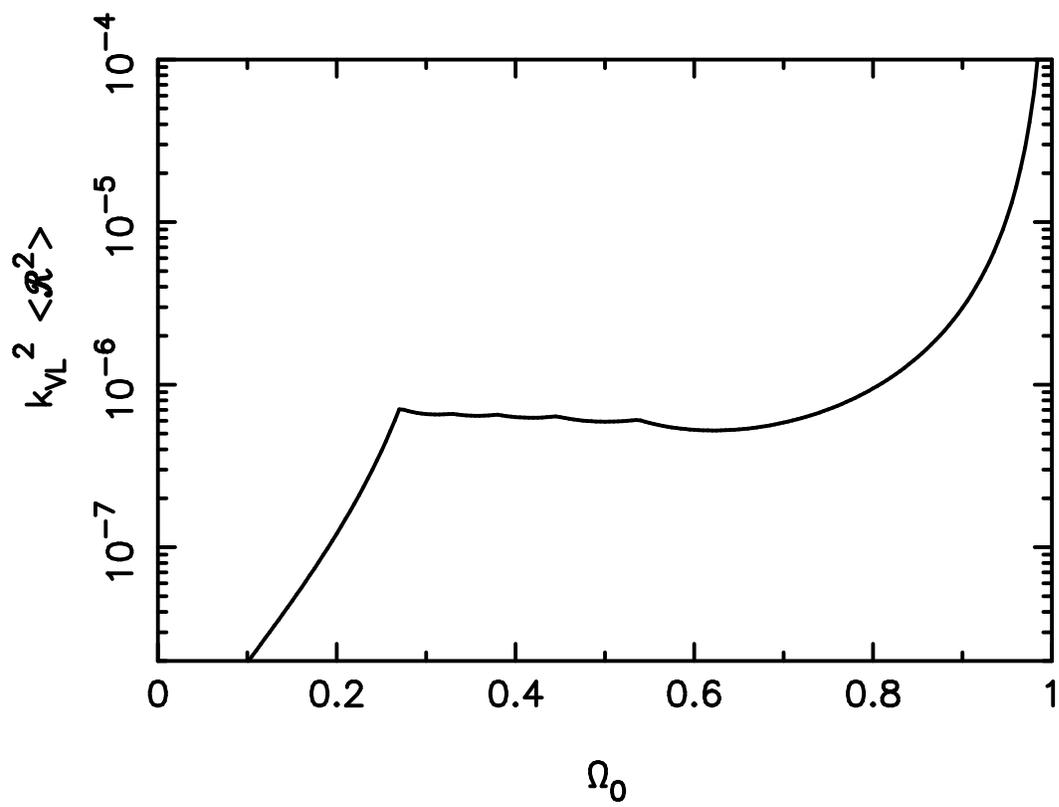
\end{document}